\documentclass[final]{IEEEtran}

\usepackage{graphicx}  
\usepackage{amsthm}
\usepackage{amssymb}
\usepackage{tikz}
\usetikzlibrary{positioning,calc}
\usepackage{amsmath}
\usepackage{comment}
\usepackage{url}
\usepackage{hyperref}

\newcommand{\T}{\mathsf{T}}

\newcommand{\diag}{\operatorname{diag}}
\newcommand{\st}{\, | \,}
\newcommand{\be}{\begin{equation}}
\newcommand{\ee}{\end{equation}}
\newcommand{\R}{\mathbb R}

\newtheorem{Theorem}{Theorem}

\newtheorem{Lemma}{Lemma}
\newtheorem{Corollary}{Corollary}

\newtheorem{Remark} {Remark}

\newcommand{\such}{\, | \,}

\title{$k$-Contraction in a Generalized  Lurie System}
\date{March 2023}

\author{Ron Ofir,  
 Jean-Jacques Slotine, and Michael Margaliot
\thanks{RO is with  the School of Elec.  Eng., Tel-Aviv University, Israel. His research is partially supported by a Khazanov post-doctoral scholarship.  JJS is with the Dept. of Mechanical Eng. and
        the Dept.  of Brain and Cognitive Sciences, Massachusetts Institute of Technology, Cambridge, Massachusetts, USA. MM (Corresponding Author) is  with the School of Elec.  Eng.,
		and the Sagol School of Neuroscience, 
		Tel-Aviv University, Tel-Aviv~69978, Israel.
		E-mail: \texttt{michaelm@tauex.tau.ac.il}}%
}

\begin{document}

\maketitle

\begin{abstract}
We derive a  sufficient  condition for~$k$-contraction in a generalized Lurie system~(GLS), that is, the feedback  connection of a nonlinear dynamical system and a memoryless nonlinear function. For~$k=1$,
this reduces to a sufficient condition  for standard contraction. 
For~$k=2$, this condition implies that every bounded solution of the  GLS converges to an equilibrium, which is not necessarily unique.
    We demonstrate the theoretical results by analyzing~$k$-contraction in
    a biochemical control circuit with nonlinear dissipation terms.
\end{abstract}


\section{Introduction}

Contraction theory~\cite{LOHMILLER1998683,bullo_contraction,sontag_cotraction_tutorial} provides a  powerful approach  for the global analysis of nonlinear systems, and has found applications  in numerous fields including   neural networks~\cite{cont_hopfield,Davydov2022NonEuclideanContract,kozachkov2022robust,kozachkov2020achieving},
robotics~\cite{chung, chuchu, illinois,singh}, learning theory~\cite{slotine_PLOS_ONE, tu, singh,manchester_eq,kozachkov2023generalization,kozachkov2022rnns,bullo_eq}, 
multi-agent systems~\cite{pham}, nonlinear control~\cite{parrilo,ccm},
systems biology~\cite{entrain2011,Russo2011,RFM_entrain},
and adaptive control~\cite{boffi,tsukamoto}.
However, if a contracting  system admits an equilibrium then this equilibrium  is   unique, and  thus  contraction theory cannot be directly applied to systems that admit  more than a single equilibrium, such as multi-stable systems that are ubiquitous in many fields of science~\cite{multi_stab_sontag,multistableNNs}. Note though that contraction theory has been extended to address 
applications such as synchronization using
the so called  virtual system~\cite{slotine_sync,pham}.

Motivated by the seminal work  of Muldowney and Li~\cite{muldowney1990compound,  li1995}  (see also~\cite{leonov}), Wu et al.
recently developed the notion of~$k$-contraction~\cite{kordercont}, with~$k>0$ an integer,
and more generally of~$\alpha$-contraction~\cite{wu2020generalization}, with~$\alpha>0$ real. Roughly speaking, a system is $k$-contracting if its flow contracts  $k$-dimensional bodies, and $\alpha$-contracting if its flow contracts bodies with Hausdorff dimension $\alpha$.

Here, we focus on $k$-contraction. For~$k=1$, 1-contraction is   standard contraction. In a time-invariant~$2$-contracting system every bounded solution converges to an equilibrium, yet the equilibrium is not necessarily unique. This allows global analysis of nonlinear systems that are not contracting (i.e., not $1$-contracting).

One of the advantages of contraction theory is that interconnections of contracting sub-systems are often contracting overall~\cite{LOHMILLER1998683,network_contractive}. However, this property does not generalize directly to $k$-contraction; for example, the series interconnection of two 2-contracting sub-systems is not 2-contracting in general~\cite{weak_contraction,Ofir2021serial}.
Refs.~\cite{ofir2021sufficient,manchester2015combination} present several sufficient conditions guaranteeing $k$-contraction in parallel and series interconnections, as well as in skew-symmetric feedback interconnections. A small-gain condition guaranteeing 2-contraction in a feedback interconnection is presented in~\cite{angeli_feed}. The main approach to proving $k$-contraction is using a concept from multilinear algebra known as $k$-compounds~\cite{comp_long_tutorial}. One reason that studying $k$-contraction in interconnections is difficult is that $k$-compounds of block matrices do not necessarily exhibit an obvious block structure~\cite{ofir2024multiplicative}. This challenge is partially addressed in~\cite{Dalin2022Duality_kcont} which derives a sufficient condition for $k$-contraction that does not require computing $k$-compounds, at the cost of potentially conservative results.
Another approach, based on  LMI-type  conditions,   has been recently suggested in~\cite{cecilia2023generalized,zoboli2023}.

Here, we consider a specific  type of feedback interconnections referred to as a \emph{generalized  Lurie system}~(GLS). 
By leveraging this special structure we obtain 
an explicit  sufficient condition for~$k$-contraction.

A Lurie system, named after  
A.  I. Lurie\footnote{Sometimes spelled Lure, Lur'e or Lurye.}~\cite{fradkov_abs_stab,LIBERZON_abs_stab},  is the feedback connection of a linear 
time-invariant~(LTI) dynamical system  and a nonlinear function. This may be interpreted as the simplest 
generalization  of the feedback connection of an LTI system and a memoryless linear controller, where the controller is replaced by a nonlinear function.
The well-known absolute stability problem is  to find conditions guaranteeing asymptotic stability of the closed-loop system for every nonlinear function in some   admissible class, e.g., sector bounded functions~\cite{khalil_book}. 

Lurie  systems were first studied as a tool to analyze the stability of surfacing submarines~\cite{LURIE200135}, but it turns out that many dynamical systems that combine linear and nonlinear dynamics
can be represented as a Lurie system. In addition, the analysis of the absolute stability problem led to many important theoretical advances in systems and control theory including the KYP Lemma and passivity-based analysis of interconnected systems~\cite{khalil_book}, 
and  the  formulation of an optimal control approach for stability analysis of switched linear systems~\cite{MARGALIOT20062059}.
 
Several authors studied the stability of Lurie systems using contraction theory (see, e.g.,~\cite{Andrieu2019LMIContract,control_syn_vincent, Proskurnikov2023GeneralizedSLemma} and also~\cite{Nguyen21} in  the  more general context of DAEs), but such systems often admit more than one equilibrium, which implies that the system is not 1-contracting. Ref.~\cite{Ron_Lurie} derived sufficient conditions for~$k$-contraction in Lurie systems and demonstrated their applications in the case~$k=2$ to several applications   
including recurrent neural networks
and opinion dynamics models.

Here, we derive new sufficient conditions for~$k$-contraction for a   more general class of systems that we refer to as a 
generalized Lurie system~(GLS). Our results generalize earlier works in two   aspects. First, we consider a wider class of systems, where both the dynamic sub-system and the memoryless sub-system are  nonlinear. This adds significant flexibility in choosing an appropriate Lurie representation for a given closed-loop system, extending the applicability of our earlier sufficient condition for $k$-contraction in~\cite{Ron_Lurie}. Second, we allow studying $k$-contraction with respect to a state-dependent metric, adding a significant  additional degree of freedom~\cite{LOHMILLER1998683}. We demonstrate our theoretical results by analyzing the global behaviour of a biochemical control circuit with nonlinear dissipation terms.

  Vectors [matrices] are denoted by small [capital] letters.
For~$x\in\R^n$ and~$1\leq p \leq \infty$, $|x|_p$ is the~$L_p$ norm of~$x$. 
 $A^\T$ is the transpose of the matrix~$A$. For a square matrix~$A$, $\det(A)$ is the determinant of~$A$. $I_n$ is the~$n\times n$ identity matrix. 
For a symmetric matrix~$A\in\R^{n\times n}$, 
the eigenvalues of~$A$ are
$
\lambda_1(A)\geq\lambda_2(A)\geq\dots\geq \lambda_n(A).
$
A symmetric matrix $A \in \R^{n \times n}$ is called positive definite [positive semi-definite], denoted~$A \succ 0$ [$A \succeq 0$], if $x^\T A x > 0$ for any~$x\in\R^n \setminus\{0\}$ [$x^\T A x \ge 0$ for any $x \in \R^n$].
For an integer~$k>0$, let~$[1,k]:=\{1,2,\dots,k\}$.
For a differentiable function~$f:\R \to \R$, $f'(s)$ is the derivative of~$f$ at~$s$.

\section{Preliminaries}\label{sec:pre}

\subsection{Compound  matrices}
Analysis of~$k$-contraction  
builds on the~$k$-compounds of the Jacobian of the vector field. 
We briefly review this topic, referring to~\cite{comp_long_tutorial} for more details and proofs.

Given~$A\in\R^{n\times m}$ and~$k\in[1,\min\{n,m\}]$, the~$k$-multiplicative compound matrix of~$A$, denoted~$A^{(k)}$, is the matrix that includes  all the~$k$-minors of~$A$,
ordered lexicographically. 
Thus,~$A^{(k)}$ is~$\binom{n}{k}\times\binom{m}{k}$.
For example,
for
$
A=\begin{bmatrix}
    a_{11} & a_{12}& a_{13}\\
    a_{21} & a_{22} &a_{23}
\end{bmatrix}
$
and~$k=2$, we have
\[
A^{(2)}=\begin{bmatrix}
   \det  \begin{bmatrix}
   a_{11}& a_{12}\\
   a_{21}&a_{22}
\end{bmatrix} &
 \det  \begin{bmatrix}
   a_{11}& a_{13}\\
   a_{21}&a_{23}
\end{bmatrix} &
\det  \begin{bmatrix}
   a_{12}& a_{13}\\
   a_{22}&a_{23}
\end{bmatrix} 
\end{bmatrix}.
\]

By definition,~$A^{(1)}=A$, and if~$A\in\R^{n\times n}$ then~$A^{(n)}=\det (A)$. Also,~$I_n^{(k)}=I_r$, with~$r:=\binom{n}{k}$, and~$(A^{(k)})^\T = (A^\T)^{(k)}$.

The Cauchy-Binet theorem asserts that for any~$A\in\R^{n\times m}$, $B\in\R^{m\times p}$ and~$k\in
[1,\min\{n,m,p\}]$, we have
\be\label{eq:cbin} 
(AB)^{(k)}=A^{(k)} B^{(k)}.
\ee
This justifies the term multiplicative compound.
For~$A,B\in\R^{n\times n}$ and~$k=n$, this reduces to~$\det(AB)=\det(A)\det(B)$.  Eq.~\eqref{eq:cbin} implies that when~$A$ is square,~$A^{(k)}$ is invertible iff~$A$ is invertible, and $(A^{(k)})^{-1} = (A^{-1})^{(k)}$.
If~$D\in\R^{n\times n}$ is diagonal, that is,~$D=\diag(\lambda_1,\dots,\lambda_n)$ then~$D^{(k)}$ is also diagonal, with
$
D^{(k)}=\diag\left(\prod_{i=1}^k\lambda_i, (\prod_{i=1}^{k-1}\lambda_i) \lambda_{k+1},\dots,
\prod_{i=n-k+1}^n\lambda_i\right).
$
More generally, 
the~$k$-multiplicative compound of a square matrix~$A\in\R^{n\times n}$ has an important  spectral property. If~$\lambda_1,\dots,\lambda_n$ are the eigenvalues of~$A$ then the eigenvalues of~$A^{(k)}$ are all the~$\binom{n}{k}$ products:
$
\lambda_{i_1}\dots\lambda_{i_k}$, $1\leq i_1<\dots< i_k\leq n$. 
It follows that if $A \succ 0$ [$A \succeq 0$]
then~$A^{(k)} \succ 0$ [$A^{(k)} \succeq 0$].

\begin{figure} 
    \centering
    \begin{tikzpicture}[scale=0.65]
        \coordinate (x1) at (3,-0.5,0);
        \coordinate (x2) at (2,2,0);
        \coordinate (x3) at (1,0,-1);
        
        \draw[dashed,fill opacity=0.5,fill=green!10] (0,0,0) -- (x1) -- ($(x1) + (x2)$) -- (x2) -- cycle;
        \draw[dashed,fill opacity=0.5,fill=green!10] (x3) -- ($(x3) + (x1)$) -- ($(x3) + (x1) + (x2)$) -- ($(x3) + (x2)$) -- cycle;

        \draw[dashed,fill opacity=0.5,fill=green!10] (0,0,0) -- (x2) -- ($(x2) + (x3)$) -- (x3) -- cycle;
        \draw[dashed,fill opacity=0.5,fill=green!10] (x1) -- ($(x1) + (x2)$) -- ($(x1) + (x2) + (x3)$) -- ($(x1) + (x3)$) -- cycle;
        
        \draw[thick,->] (0,0,0)--(x1) node[right]{$x^1$};
        \draw[thick,->] (0,0,0)--(x2) node[above]{$x^2$};
        \draw[thick,->] (0,0,0)--(x3) node[above left=-0.1cm]{$x^3$};
        \draw (0,0,0) node[below]{0};
        \draw (3.15,1,0.5) node[]{$\mathcal{P}(x^1,x^2,x^3)$};
    \end{tikzpicture}
    \caption{A 3D parallelotope.}
    \label{fig:paralellotope}
\end{figure}
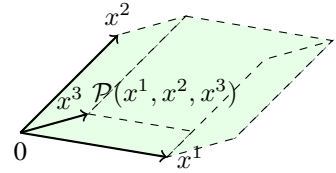

The~$k$-multiplicative compound has an important geometric interpretation. Fix~$k$ vectors~$a^1,\dots,a^k\in\R^n$. The parallelotope generated by these vectors is
$
\mathcal P(a^1,\dots,a^k):=\left\{\sum_{i=1}^k c_i  a^i\such c_i\in[0,1]\right\}
$
(see Fig.~\ref{fig:paralellotope}). Define  
$
V:=\begin{bmatrix} a^1&\dots& a^k\end{bmatrix} 
\in\R^{n\times k}$.
Then the volume of~$\mathcal P(a^1,\dots,a^k)$  is equal to
$
|V^{(k)}|_2.
$
Note that~$V^{(k)}$ has dimensions~$\binom{n}{k}\times\binom{k}{k} = \binom{n}{k}\times1$, i.e., it is a column vector. In the particular case~$k=n$, this gives the well-known expression: 
$
\text{volume} (\mathcal P(a^1,\dots,a^n))=|\det\begin{bmatrix} a^1&\dots& a^n \end{bmatrix} |.
$

To study the evolution of a parallelotope
under a dynamical system, consider the  LTI system
\be\label{eq:LTI}
\dot x=Ax
\ee
and let~$x(t,x_0)$ denote its solution at time~$t$ when~$x(0)=x_0$. Fix initial conditions~$a^1,\dots,a^k\in\R^n$, and consider
the time-varying 
parallelotope~$
\mathcal P(x(t,a^1),\dots,x(t,a^k)).
$
We already know that the volume of this 
parallelotope is the~$L_2$ norm of the  vector 
$
v(t):=\begin{bmatrix} x(t,a^1)&\dots& x(t,a^k) \end{bmatrix} ^{(k)}
$. Now,
\begin{align}\label{eq:ddtvt}
    \dot v(t)&= \frac{d}{dt} \big (\exp(At) \begin{bmatrix} a^1&\dots& a^k \end{bmatrix} \big )^{(k)}\nonumber\\
    &=\frac{d}{dt} \left ((\exp(At)) ^{(k)}\begin{bmatrix} a^1&\dots& a^k \end{bmatrix} ^{(k)}\right )\nonumber \\
    &=\frac{d}{dt} \left ((\exp(At) )^{(k)}\right ) v(0)  .
\end{align}

Eq.~\eqref{eq:ddtvt} leads  to the notion of the $k$-additive compound of a (square) matrix~$A\in\R^{n\times n}$. For~$\varepsilon\geq 0$, consider the~$\binom{n}{k}\times\binom{n}{k}$ matrix~$(I_n+\varepsilon A)^{(k)}$. Every entry of this matrix is a~$k$-minor of~$I_n+\varepsilon A$, so 
$
(I_n+\varepsilon A)^{(k)}=B_0\varepsilon^0+B_1\varepsilon^1+\dots+B_k \varepsilon^k.
$
Setting~$\varepsilon=0$ gives~$B_0=I_r$, with~$r:=\binom{n}{k}$.
The matrix~$B_1$ is called the~$k$-additive compound of~$A$, denoted~$A^{[k]}$. Thus,
\[
\frac{d}{dt} (\exp(At))^{(k)} |_{t=0}=\frac{d}{dt} (I_n+A t)^{(k)} |_{t=0} = A^{[k]},
\]
and~\eqref{eq:ddtvt} yields  
$
    \dot v(t) = A^{[k]} v(t)  .
$
It follows from the Cauchy-Binet theorem that for any~$A,B\in\R^{n\times n}$, we have~$(A + B)^{[k]} = A^{[k]} + B^{[k]}$, and hence the term  additive compound. Furthermore, it follows from the definitions above  that~$(A^{[k]})^\T = (A^\T)^{[k]}$, and that if~$T\in\R^{n\times n}$ is non-singular then
$
(TAT^{-1})^{[k]} = T^{(k)} A^{[k]}  (T^{(k)})^{-1}.
$

The~$k$-additive compound of~$A\in\R^{n\times n}$ has an important  spectral property. If~$\lambda_1,\dots,\lambda_n$ are the eigenvalues of~$A$ then the eigenvalues of~$A^{[k]}$ are all the~$\binom{n}{k}$ sums:
$
\lambda_{i_1}+\dots+\lambda_{i_k}$, $1\leq i_1<\dots< i_k\leq n. 
$
In particular,~$A^{[k]}$ is Hurwitz  iff the sum of every $k$ eigenvalues of~$A $ has a negative real part, and then
the volume of every~$k$-parallelotope decays to zero exponentially under~\eqref{eq:LTI}.

\subsection{$k$-contraction w.r.t. a Riemannian metric}
Consider the non-linear time-varying system
\begin{equation}\label{eq:nonlin_tv}
    \dot x(t) = f(t,x),
\end{equation}
with~$f:\R_{\geq0} \times \R^n\to \R^n$, evolving on a convex state-space~$\Omega\subseteq\R^n$. Let~$J(x) := \frac{\partial}{\partial x}f(x)$ denote the Jacobian of~$f$. Fix~$k \in [1,n]$. Let~$\mathcal S^k := \{r \in \R^k_{\geq 0} \st  \sum_{i=1}^k r_i \le 1\}$ denote the unit simplex in $\R^k$. Fix arbitrary initial conditions~$a^1,\dots,a^{k+1}\in\Omega$, and define~$h  : \mathcal S^k \to \R^n$ by
\begin{equation*}
    h(r) := \sum_{i=1}^k r_i a^i + \left(1 - \sum_{i=1}^k r_i\right) a^{k+1}.
\end{equation*}
Let~$w^i(t,r) := \frac{\partial}{\partial r_i} x(t,h(r)) $, 
i.e., the difference in the solution at time~$t$ corresponding to  a small change  in the initial condition~$h(r)$ by modifying~$r_i$.
Let
$   W(t,r) := \begin{bmatrix}
        w^1(t,r) & \dots & w^k(t,r)
    \end{bmatrix}.
$
Then~$\frac{d}{dt} W(t,r) = J(t,x(t,h(r))) W(t,r)$, 
i.e.,~$W(t,r)$ satisfies an  LTV dynamics, and thus
$
    W^{(k)}(t,r) = J^{[k]}(t,x(t,h(r))) W^{(k)}(t,h(r)).
$

System~\eqref{eq:nonlin_tv}
is said to be $k$-contracting~\cite{kordercont}   if~$
    |W^{(k)}(t,r)|$
    decays  exponentially with~$t$ for any~$r\in\mathcal S^k$. 
  A sufficient condition for this   is that 
$
    \mu(J^{[k]}(t,x)) \le -\eta<0$,  for all~$x\in\Omega,t\geq0$,
where $\mu$ is  the matrix measure induced by the norm~$|\cdot|$. 

Let~$\Theta  : \Omega \to \R^{n \times n}$ be~$C^1$, satisfying 
$
0 \prec \sigma_1 I_n \preceq \Theta^\T(x) \Theta(x) \preceq \sigma_2 I_n
$
  for   all $x \in \Omega$ for some~$\sigma_2 \ge \sigma_1 > 0$.
  Let~$\ell:=\binom{n}{k}$. Then 
  \be\label{eq:s1s2}
0 \prec \sigma_1^k I_\ell \preceq (\Theta^\T(x) \Theta(x))^{(k)} \preceq \sigma_2^k I_\ell \text{ for all } x\in\Omega.
\ee

Let~$\tilde W^{(k)}(t,r):=\Theta^{(k)}(x(t,h(r))) W^{(k)}(t,r)
$.
Our goal is to establish contraction of~$ 
|\tilde W^{(k)}(t,r)|_2$, i.e., the scaled~$L_2$ norm of~$W^{(k)}$.

\begin{Lemma}\label{lem:k_cont_scaled}
Let~$\dot \Theta : \Omega  \to \R^{n \times n}$  denote the matrix whose~$(i,j)$ entry is 
$
    \left(\dot \theta(x)\right)_{ij} := \left(\frac{\partial \theta_{ij}(x)}{\partial x}\right)^\T f(t,x).
$ 
Suppose that
\begin{equation*}
    \mu_2 \big ( (  \dot \Theta(x) \Theta^{-1} (x)+\Theta(x) J(t,x) \Theta^{-1}(x)  )^{[k]}   \big ) \le -\eta < 0,
\end{equation*}
for all $t\geq 0,x\in\Omega$. Then    for any $t \geq 0,r\in\mathcal{S}^k$, we have 
\begin{equation}\label{eq:Thetakcont}
  \left  |\tilde  W^{(k)}(t,r)\right |_2 \le \exp(-\eta t) \left |\tilde  W^{(k)}(0,r)\right |_2 .
\end{equation}
\end{Lemma}

\begin{IEEEproof}
Using the chain rule (omitting for simplicity the arguments 
in~$\Theta,W$) gives
\begin{align*}
    \frac{d}{dt}  \big (\Theta  W \big) &= \left(\dot \Theta  + \Theta  J \right) W   
    &= \left(\dot \Theta  \Theta^{-1}  + \Theta  J  \Theta^{-1} \right) \Theta  W ,
\end{align*}
and this completes the proof.
\end{IEEEproof}

 Note that combining~\eqref{eq:Thetakcont}  with~\eqref{eq:s1s2} implies that     
\begin{equation} 
    |  W^{(k)}(t,r)|_2 \le
    ( \sigma_2/\sigma_1 )^{1/2} 
    \exp(-\eta t) |  W^{(k)}(0,r)|_2 , 
\end{equation}
for all~$t\geq 0$ and~$r\in \mathcal S^k$.
This implies that~\eqref{eq:nonlin_tv} satisfies all the asymptotic properties of $k$-contracting systems. 

\subsection{Singular values of the product of two matrices}
Let~$\sigma_1(A)\geq\sigma_2(A)\geq\dots$ denote the ordered singular values of the matrix~$A$. We require the following result.
\begin{Lemma}\label{lemmma:sing}\cite[Thm.~3.3.14]{Horn1991TopicsMatrixAna}
Fix~$A\in\R^{m\times p},B\in\R^{p\times n}$. For any $k\in[1,\min\{m,p,n\}]$, and~$s>0$, we have 
$    \sum_{i=1}^k \sigma_i^s(AB) \leq
    \sum_{i=1}^k (\sigma_i(A)\sigma_i(B))^s  .
$
 \end{Lemma}

\section{Generalized Lurie systems}\label{II}
We now describe the~GLS that is analyzed in this note.
We first recall the standard Lurie system. 
Consider  the  feedback interconnection of  
a  linear system
\begin{align}\label{eq:yeslin}
\dot x =Ax+Bu, \quad 
y =C x,
\end{align}
where $x\in\R^n$ is the state, $u\in\R^m$ is the input,~$y\in\R^p$ is the output,
with a memoryless nonlinear function~$\Phi:\R^p\to\R^m$, that is,
\be\label{eq:Phi}
u=-\Phi(y). 
\ee
The closed-loop system~\eqref{eq:yeslin}-\eqref{eq:Phi}  is called a Lurie system.

Stability analysis of Lurie systems plays an important role in systems and control theory (see, e.g.,~\cite{vidyasagar2002nonlinear, khalil_book}), and has found applications  in power systems~\cite{power_syst_lurie},  continuous time Hopfield neural networks, and as general models for learning and system identification~\cite{Man_learning}. Hopfield networks, and many   other examples, can be written as the general networked system
$
\dot x=-Dx +W_1 h (W_2x),
$
with~$D\in\R^{n\times n}$, $h:\R^s\to \R^q$, and~$W_1\in \R^{n\times q}$, $W_2 \in \R^{    s \times n}$. This can be represented in the form
\begin{align}\label{eq:lui}
    \dot x&= - D  x +   u,\quad
    y=x,\nonumber\\
    u&=   W_1 h(W_2  y  ) ,
\end{align}
which is a Lurie system.

A Lurie system typically admits more than a  single equilibrium point,  and thus it is not contracting w.r.t. any norm. However, it may still be~$k$-contracting, with~$k>1$. In particular, if the system is~$2$-contracting then,    since the closed-loop system is time-invariant,      every bounded solution of the system converges to an equilibrium (which is not necessarily unique),  see, e.g.,~\cite{li1995,kordercont}.
 The recent paper~\cite{Ron_Lurie} analyzed $k$-contraction in Lurie systems.

\subsection{Generalized Lurie system}

\begin{figure}
    \centering
    \begin{tikzpicture}[scale=0.5,
        block/.style = {draw, rectangle, thick, minimum height=2em, minimum width=3em},
        sum/.style = {draw, circle, thick, minimum width=2em}]
        
        \node[block, minimum height=1.0cm, minimum width=2.4cm] (LTI) {$\begin{aligned}\dot{x} &= f(x,u) \\ y &= g(x)\end{aligned}$} ;
        \node[block, minimum height=1cm, minimum width=1cm] (nonlin) [below=0.3cm of LTI] {$\Phi$} ;
        \node[sum, left=1cm of LTI] (fbsum) {};
        
        \draw[->, thick] (LTI.east) -- node[above, midway] {$y $} +(1,0) coordinate(LTIaux) |- (nonlin.east);
        \draw[->, thick] (LTIaux) -- +(1,0);
        \draw[->, thick] (nonlin.west) -| (fbsum.south) node[below left] {$-$};
        \draw[->, thick] (fbsum.east) -- (LTI.west) node[midway,above] {$u $};
        \draw[<-, thick] (fbsum.west) node[above left] {$+$} -- +(-1,0) node[left] {$0$};
        \end{tikzpicture}
    \caption{Block diagram of a generalized Lurie system.}
    \label{fig:nonlin_lurie}
\end{figure}
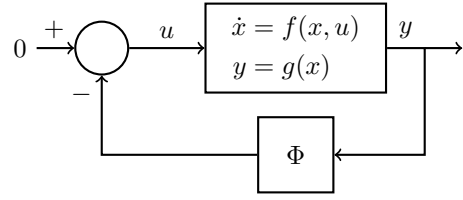

Consider  the  feedback interconnection of  a nonlinear  dynamical   system
\begin{align}\label{eq:nonlin}
\dot x =f(x,u), \quad 
y =g(x), 
\end{align}
where~$x,u,y$ are as in~\eqref{eq:yeslin},  
  with a memoryless nonlinear function~$\Phi$
as in~\eqref{eq:Phi}
(see Fig.~\ref{fig:nonlin_lurie}).
We assume that~$f,g$ and~$\Phi$ are~$C^1$.
%
Combining~\eqref{eq:nonlin}
and~\eqref{eq:Phi}
yields the closed-loop system  
\begin{equation}\label{eq:cl}
    \dot x = f_{cl}(x) :=   f(x,-\Phi(g(x))). 
\end{equation}
The  Jacobian~$J_{cl}(x) := \ \frac{\partial }{\partial x} f_{cl} (x)$ of this
closed-loop system is equal to 
\begin{align*}
\frac{\partial f}{\partial x}(x,-\Phi(g(x)))
 - \frac{\partial f}{\partial u}(x,-\Phi(g(x))) \frac{\partial \Phi}{\partial y}(g(x)) \frac{\partial g}{\partial x}(x) . 
\end{align*}
We refer to this system as a~GLS. 
Ref.~\cite{angeli_feed} derived a sufficient condition for~$2$-contraction  of a general feedback system, but the special structure of a GLS allows for more explicit results.

\section{Main result}\label{sec:main}
Our main result 
provides  a sufficient condition for $k$-contraction of a GLS
w.r.t. a  space-dependent scaled~$L_2$ metric. For~$k=1$, this reduces to a sufficient condition for standard contraction.

Let~$\Theta:\R^n \to \R^{n \times n}$ be $C^1$, and such that~$0 \prec \sigma_1 I_n \preceq \Theta^\T(x)  \Theta(x) \preceq \sigma_2 I_n$ for all~$x\in\R^n$. Define the derivatives of $\Theta$ along solutions of the open-loop and closed-loop systems~$\dot \Theta_{ol}:\R^n \times \R^m \to \R^{n \times n}$ and~$\dot \Theta_{cl}:\R^n \to \R^{n \times n}$ by
\begin{align*}
    \left(\dot \theta_{ol}(x,u)\right)_{ij} &:= \left(\frac{\partial \theta_{ij}}{\partial x}(x)\right)^\T f(x,u),   \\
    \left(\dot \theta_{cl}(x)\right)_{ij} &:= \left(\frac{\partial \theta_{ij}}{\partial x}(x)\right)^\T f_{cl}(x),
\end{align*}
respectively. 
Note that  
\begin{align*}
    \left(\dot \theta_{ol}(x, -\Phi(g(x)))\right)_{ij} &= \left(\frac{\partial \theta_{ij}}{\partial x}(x)\right)^\T f(x, -\Phi(g(x))) \\
        &= \left(\dot \theta_{cl}(x)\right)_{ij}.
\end{align*}
Define the Riemannian Jacobians~\cite{LOHMILLER1998683} for the open-loop  and closed-loop systems by:
\begin{align}
    \tilde{J}_{ol}(x,u) &:= \Theta(x)\frac{\partial f}{\partial x}(x,u) \Theta^{-1}(x) + \dot \Theta_{ol}(x) \Theta^{-1}(x),  \label{eq:tildeJol}\\
	\tilde{J}_{cl}(x) &:= \Theta(x)J_{cl}(x)\Theta^{-1}(x) + \dot \Theta_{cl}(x) \Theta^{-1}(x). \label{eq:tildeJcl}
   \end{align} 
  We can now state our main result. For the sake of brevity, we write~$f$ for~$f(x,u)$,  $\Theta$ for~$\Theta(x)$, and so on. 

\begin{Theorem}\label{thm:suff}
Define the symmetric matrix
\begin{align}\label{eq:mat_H}
H&:=\tilde J_{ol}  + \tilde J_{ol}^\T
         + \Theta  \frac{\partial f}{\partial u}   \frac{\partial f ^\T}{\partial u}  \Theta^\T
         + \Theta^{-\T}  \frac{\partial g ^\T}{\partial x}  \frac{\partial g}{\partial x} \Theta^{-1}. 
\end{align}
	Fix~$k \in [1,n]$.
    Assume  that there exist $\eta_1,\eta_2 \in \R$ such that
	\begin{align}\label{cond:diff_ARI_eig}
		\sum_{i=1}^k \lambda_i  (H ) \le -\eta_1, \text{ for all } x,u,
	\end{align}
  and also that at least one of the following two conditions holds: either
	\begin{equation}\label{cond:B_JPhi}
		\sum_{i=1}^k \lambda_i\left(\Theta  \frac{\partial f}{\partial u} \left(\frac{\partial \Phi}{\partial y}\frac{\partial \Phi}{\partial y}^\T  - I_m\right)\frac{\partial f}{\partial u} ^\T \Theta^\T \right) \le -\eta_2
	\end{equation}
    or
    \begin{equation}\label{cond:C_JPhi}
    	\sum_{i=1}^k \lambda_i\left(\Theta^{-\T} \frac{\partial g}{\partial x}^\T\left(\frac{\partial \Phi}{\partial y}^\T\frac{\partial \Phi}{\partial y}- I_p\right)\frac{\partial g}{\partial x} \Theta^{-1}\right) \le -\eta_2
    \end{equation}
	for all $x,u$.
 Then
	\begin{equation}
		\sum_{i=1}^k\lambda_i\left(\tilde{J}_{cl}(x) + \tilde{J}_{cl}^\T(x)\right) \le -\eta_1 - \eta_2 \text{ for all } x.
	\end{equation}
	  In particular, if~$\eta_1+\eta_2>0$ then the closed-loop system~\eqref{eq:cl} 
 is~$k$-contracting  w.r.t. the scaled~$L_2$ norm~$z\to|\Theta(z) z|_2$ with rate~$(\eta_1+\eta_2)/2$.
\end{Theorem}

\begin{IEEEproof}
To simplify  the notation, let
    \begin{alignat*}{2}
            \tilde A(x,u) &:= \Theta(x)\frac{\partial f}{\partial x}(x,u)\Theta^{-1}(x), & \; 
            \tilde B(x,u) &:= \Theta(x) \frac{\partial f}{\partial u}(x,u), \\
            \tilde C(x) &:= \frac{\partial g}{\partial x}(x)\Theta^{-1}(x), &
            J_\Phi(y) &:= \frac{\partial \Phi}{\partial y}(y).
    \end{alignat*}
    By~\eqref{eq:tildeJcl},
	\begin{align*}
		\tilde J_{cl}(x) &= \Theta(x)J_{cl}(x)\Theta^{-1}(x) + \dot \Theta_{cl}(x) \Theta^{-1}(x) \\
            &= \tilde A(x,-\Phi(g(x))) + \dot \Theta_{cl}(x) \Theta^{-1}(x) \\ &- \tilde B(x,-\Phi(g(x))) J_\Phi(g(x)) \tilde C(x).
    \end{align*}
    Omitting the arguments of the Jacobians and of~$P$, and using~\eqref{cond:diff_ARI_eig} yields
    that~$(\tilde J_{cl} + \tilde J_{cl}^\T)^{[k]}$ is equal to
    \begin{align}\label{eq:tildeJbound}
    	  & \left(\tilde A + \tilde A^\T + \dot \Theta_{cl} \Theta^{-1} + \Theta^{-\T} \dot \Theta_{cl}^\T - \tilde B J_\Phi \tilde C - \tilde C^\T J_\Phi^\T \tilde B^\T \right)^{[k]} \nonumber \\
    		&\preceq -\eta_1 I_r - \left(\tilde B \tilde B^\T + \tilde C^\T \tilde C + \tilde B J_\Phi \tilde C + \tilde C^\T J_\Phi^\T \tilde B^\T \right)^{[k]} \\
    		&= -\eta_1 I_r - \left(\tilde B \tilde B^\T + \tilde C^\T \tilde C + (\tilde B + \tilde C^\T J_\Phi^\T)(\tilde B + \tilde C^\T J_\Phi^\T)^\T \right. \nonumber \\
            &\quad\quad\quad\quad\quad \left. - \tilde B \tilde B^\T - \tilde C^\T J_\Phi^\T J_\Phi \tilde C \right)^{[k]} \nonumber \\
    		&\preceq -\eta_1 I_r + \left(\tilde C^\T (J_\Phi^\T J_\Phi - I_p) \tilde C \right)^{[k]} \nonumber . 
    \end{align}
Now if~\eqref{cond:C_JPhi} holds then  
    \be\label{eq:eta1eta2}
    	(\tilde J_{cl}  + \tilde J_{cl}^\T )^{[k]} 
    		\preceq -(\eta_1 + \eta_2) I_r . 
\ee
 Alternatively, continuing similarly from~\eqref{eq:tildeJbound}, it can also be shown that
$
    	(\tilde J_{cl}  + \tilde J_{cl}^\T )^{[k]}  \preceq -\eta_1 I_r + \left(\tilde B (J_\Phi J_\Phi^\T - I_m) \tilde B^\T \right)^{[k]}  $, 
 and condition~\eqref{cond:B_JPhi} implies that~\eqref{eq:eta1eta2} holds. 
       Applying Lemma~\ref{lem:k_cont_scaled} completes the proof of Thm.~\ref{thm:suff}. 
\end{IEEEproof}

Note that condition~\eqref{cond:diff_ARI_eig}
only involves~$f$ and~$g$, that is, the nonlinear dynamical system.
Conditions~\eqref{cond:B_JPhi} and~\eqref{cond:C_JPhi} also involve  the nonlinear 
function~$\Phi$. Note also that there  is no requirement on the individual signs of~$\eta_1$ and~$\eta_2$, but only on the sign of their sum. 

  Thm.~\ref{thm:suff} is stated 
with a state-dependent metric, as this provides more generality and there exist numerical algorithms for searching for a suitable~$\Theta(x)$  in the case of~$1$-contraction~\cite{parrilo,tu,ccm}.

\begin{Remark}\label{rem:diff_ARI}
When~$\Theta$ is  constant, condition~\eqref{cond:diff_ARI_eig} can be written in a form reminiscent of an algebraic Riccati inequality~(ARI) \cite{LMIs}. To show this, Let~$P : = \Theta^\T \Theta$. Eq.~\eqref{cond:diff_ARI_eig} is equivalent to 
    \begin{align*}
       & \left( \tilde J_{ol}+  \tilde J_{ol} ^\T   
         + \Theta \frac{\partial f}{\partial u}  \frac{\partial f ^\T}{\partial u} \Theta^\T + \Theta^{-\T} \frac{\partial g ^\T}{\partial x} \frac{\partial g}{\partial x}\Theta^{-1} \right)^{[k]}
         \preceq -\eta_1 I_r.
    \end{align*}
Multiplying on the left [right] by~$\left(\Theta^\T\right)^{(k)}$ [$\Theta^{(k)}$]  gives
    \begin{align}\label{cond:diff_ARI}
     &   P^{(k)} \frac{\partial f}{\partial x}^{[k]} + \left(\frac{\partial f }{\partial x}^{[k]}\right)^\T P^{(k)}
     \nonumber\\&+ \left(\Theta^\T\right)^{(k)}\left(\Theta\frac{\partial f}{\partial u} \frac{\partial f ^\T}{\partial u} \Theta^{\T} + \Theta^{-\T} \frac{\partial g ^\T}{\partial x} \frac{\partial g}{\partial x}\Theta^{-1} \right)^{[k]}\Theta^{(k)} \nonumber\\&\preceq -\eta_1 P^{(k)}.
    \end{align}
    In particular, for $k=1$ we have the standard ARI
   $
        P \frac{\partial f}{\partial x} +  \frac{\partial f ^\T}{\partial x}  P + P\frac{\partial f}{\partial u} \frac{\partial f ^\T}{\partial u}  P +  \frac{\partial g ^\T}{\partial x} \frac{\partial g}{\partial x} \preceq -\eta_1 P$.
\end{Remark}

 \begin{Remark}
    The sufficient condition for $k$-contraction for a standard   Lurie system  in~\cite{Ron_Lurie} can be recovered as a special case of Thm.~\ref{thm:suff} when~$\Theta$ is chosen as constant and symmetric, and in addition the nonlinear dynamical sub-system~\eqref{eq:nonlin}  reduces to   an LTI, i.e., $f(x,u) = Ax + Bu$ and $g(x) = C  x$. As in Remark~\ref{rem:diff_ARI}, in this case~\eqref{cond:diff_ARI_eig} can be written as
    \begin{align}\label{cond:diff_ARI_LTI}
       & P^{(k)} A^{[k]} + (A^{[k]})^\T P^{(k)} \\&+ \Theta^{(k)}\left(\Theta BB^\T \Theta + \Theta^{-1}C^\T C\Theta^{-1} \right)^{[k]}\Theta^{(k)} \preceq -\eta_1 P^{(k)}, \nonumber
    \end{align}
    whereas~\eqref{cond:B_JPhi} and~\eqref{cond:C_JPhi} become   either 
    \begin{align*}
        \sum_{i=1}^k \lambda_i\left(\Theta B \left(\frac{\partial \Phi}{\partial y}\frac{\partial \Phi}{\partial y}^\T  - I_m\right)B^\T \Theta \right) \le -\eta_2, 
 \end{align*}
 or
   \begin{align*}
    	\sum_{i=1}^k \lambda_i\left(\Theta^{-1} C^\T\left(\frac{\partial \Phi}{\partial y}^\T\frac{\partial \Phi}{\partial y}- I_p\right)C \Theta^{-1}\right) \le -\eta_2 . 
    \end{align*}
\end{Remark}

\begin{Remark}\label{rem:C_is_I}
    There are  several  special cases 
where   conditions~\eqref{cond:B_JPhi}, \eqref{cond:C_JPhi} simplify. For example, take~$\Theta(x)=pI_n$, with~$p>0$. If~$f(x,u)=h(x)+u$
then~\eqref{cond:B_JPhi}
becomes
\begin{equation}	p^2\sum_{i=1}^k \lambda_i  \left(\frac{\partial \Phi}{\partial y}\frac{\partial \Phi}{\partial y}^\T  - I_m\right)   \le -\eta_2,
	\end{equation}
that is,
\begin{equation}	
p^2\left( -k+\sum_{i=1}^k \sigma_i^2  \left(\frac{\partial \Phi} {\partial y}    \right)  \right)   \le -\eta_2,
	\end{equation}
where~$\sigma_1(A)\geq\sigma_2(A)\geq\dots$ are
the ordered singular values of the matrix~$A$. 
    Similarly, if~$g(x)=x$ then 
\eqref{cond:C_JPhi} becomes   
    \begin{equation}\label{eq:pm2}
    	p^{-2}\left(-k+\sum_{i=1}^k \sigma^2_i  \left(\frac{\partial \Phi}{\partial y} \right)  \right) \le -\eta_2. 
    \end{equation}
\end{Remark}


\section{Applications}\label{sec:appli}

\subsection{$k$-contraction w.r.t. a state-dependent metric for tridiagonal~systems}
One possible approach for determining a suitable metric~$\Theta(x)$, based on considering~\eqref{cond:diff_ARI_eig} and~\eqref{eq:tildeJol}, is to find a~$\Theta(x)$ such that~$\Theta \frac{\partial f}{\partial x}\Theta^{-1}$ is skew-symmetric. We now demonstrate an application of this idea  to  a specific class of dynamical systems. Let~$\xi^1,\dots,\xi^n$ denote the canonical basis in~$\R^n$. 

Consider the system~\eqref{eq:nonlin} with
\[
f(x,u)=\begin{bmatrix}
    h_1(x_1,x_2)\\
    h_2(x_1,x_2,x_3)\\
    \vdots\\
    h_{n-1}(x_{n-2},x_{n-1},x_n) \\
    h_n(x_{n-1},x_n)  
\end{bmatrix} + \xi^m u,
\]
and~$y=  (\xi^\ell)^\top  x$, for some~$m,\ell\in[1,n]$. 
The vector field here corresponds for  example to  a 
set of agents ordered on a 1D chain such that every agent directly  communicates only with its two nearest neighbors. 
This is accompanied by a nonlinear static feedback~$\Phi(y)$ from~$x_\ell$ to~$\dot x_m$. 

Assume that every~$h_i$ is~$C^2$,  and let~$h_{i,j} : =\frac{\partial}{\partial x_j}h_i$. Assume also  
that there exist~$s_1,s_2>0$ such that
\begin{align}\label{eq:tri_s1_s2}
    s_1< - h_{i,i+1} <s_2 \text { and } 
    s_1<h_{i+1,i} < s_2 \text{ for all } i,x.
\end{align}
In particular, $\dot x_i$ increases [decreases] when~$x_{i-1}$ [$x_{i+1}$] increases. 
  The Jacobian~$\frac{\partial f}{\partial x}$ is the tridiagonal matrix 
     \[
    \begin{bmatrix}  h_{1,1} & h_{1,2} & 0 &0&\dots &0 &0&0 \\
     h_{2,1}&h_{2,2}&h_{2,3}&0 &\dots & 0 &0&0\\
    &&&& \vdots\\ 
     0&0&0&0&\dots&
     h_{n-1,n-2} &h_{n-1,n-1}&h_{n-1,n}\\
      0&0&0&0&\dots&0& h_{n,n-1}&h_{n,n}
     \end{bmatrix}. 
     \]
   Define
\be\label{eq:def_Theta}
\Theta(x):=\diag(  \delta_1(x),\dots,\delta_n(x)   )
\ee
with~$\delta_1(x)=1$, $\delta_2(x)=\sqrt{-h_{1,2}(x)/h_{2,1}(x)}$, $\delta_3(x)=\delta_2(x) \sqrt{ -h_{2,3}(x)/h_{3,2 }(x) }$, and so on. Then
$
M(x):=\Theta (x)\frac{\partial }{\partial x} f(x) \Theta^{-1} (x)$ is the sum of a diagonal and a skew-symmetric matrix,
so~$M+M^T=2\diag(h_{1,1},\dots,h_{n,n}   )$. 
Since~$\Theta(x)$ is diagonal, so is~$\dot \Theta_{ol}(x) \Theta^{-1}(x)$. 
Eq.~\eqref{eq:tildeJol}  implies that~$
  \tilde{J}_{ol}+  \tilde{J}_{ol}^\top $ is  diagonal, with
 \[
 (\tilde{J}_{ol}+\tilde{J}_{ol}^\top )_{i,i }  = 2h_{i,i}+ 
 \left  (   \left(\frac{\partial}{\partial x} \delta_i\right) ^\top  f    +    f^\top   \left(\frac{\partial}{\partial x} \delta_i\right) \right )\delta^{-1}_i . 
  \]
  Also,~$\Theta  \frac{\partial f }{\partial u}   \frac{\partial f  ^\T}{\partial u}  \Theta^\T$ is diagonal with all entries zero except for~$\delta_m^2$ at entry~$(m,m)$, and~$ \Theta^{-\T}   \frac{\partial g  ^\T}{\partial x}  \frac{\partial g }{\partial x} \Theta^{-1}$
 is diagonal with all entries zero except for~$\delta_\ell^{-2}$ at entry~$(\ell,\ell)$. Thus, in this case~$H$ in~\eqref{eq:mat_H} is \emph{diagonal}, and determining the value~$\eta_1$ such that~\eqref{cond:diff_ARI_eig} holds is greatly  simplified. 

Also, conditions~\ref{cond:B_JPhi}
and~\eqref{cond:C_JPhi} simplify to 
$
	\delta_m^2  ((\Phi')^{2} -1)
   \le -\eta_2
$
 and
 $
   \delta_
   \ell^{-2}  	((\Phi')^{2} -1)   \le -\eta_2$, 
	respectively.

Summarizing, when the nonlinear sub-system in the~GLS is a tridiagonal system satisfying~\eqref{eq:tri_s1_s2}  then there exists a natural state-dependent metric for analyzing~$k$-contraction.

\subsection{Networked systems}
We   describe an application of Thm.~\ref{thm:suff} to the  general nonlinear networked system:
\begin{equation}\label{eq:net_sys}
    \dot x(t) = -d(  x (t))+ W_1f \left (W_2x(t) \right )+v,
\end{equation}
where
  $x \in \Omega \subseteq \R^n$,  $d(x)=\begin{bmatrix} d_1(x_1)&\dots&d_n(x_n) \end{bmatrix}^\T$, 
  $W_1 \in \R^{n \times m}, W_2 \in \R^{q \times n} $ are matrices of interconnection weights, $v\in\R^n$ is a constant ``offset'' vector, and $f : \R^q \to \R^m$. We assume that the state space~$\Omega$ is convex,  and that~$f$ and~$d$ are~$C^1$.
    Let   
$   
    J_f(z) :  = \begin{bmatrix}
        \frac{\partial f_1}{\partial z_1}(z) & \dots & \frac{\partial f_1}{\partial z_q}(z) \\
        \vdots & \ddots & \\
        \frac{\partial f_m}{\partial z_1}(z) & \dots & \frac{\partial f_m}{\partial z_q}(z)
    \end{bmatrix} 
$  
 denote the Jacobian of the vector field~$f$. 
 
Note that~\eqref{eq:net_sys} cannot be expressed naturally as a standard Lurie system, as it does not include a linear sub-system.

Equations of the form~\eqref{eq:net_sys} are common in recurrent neural network models. 
In this context, the function~$f$ in~\eqref{eq:net_sys} is typically diagonal, that is,~$q=m$ and
  $
f(z)=\begin{bmatrix} f_1(z_1)&\dots &f_q (z_q)\end{bmatrix}^\T,
  $
  where the~$f_i$s are the neuron activation functions. Thus,~$J_f(z)$ is diagonal.
  More generally, the~$f_i$s may represent functions that are bounded or saturated, 
   and are thus nonlinear. 
 
Typically,~$d(x)$ represents nonlinear  dissipation and this will be used to establish~$k$-contraction. 
Intuitively, increasing   the dissipation   should make
the system   ``more stable''. This is formalized in
the next result.
\begin{Theorem} \label{thm:net_k_contract}
    Consider system~\eqref{eq:net_sys}. Fix~$k \in [1,n]$. Define 
    \be\label{eq:alphak}
        \alpha_k:= k^{-1} \inf \big\{ \sum_{j=1}^k d'_{i_j}(x_{i_j})  \st x\in\Omega, 1\leq i_1<\dots<i_k\leq n \big \}.
    \ee
    If $ \alpha_k>0 $ and 
    \begin{equation}\label{cond:net_k_smallgain}
         {\sup_{x \in \Omega}} \|J_f(W_2x)\|_2^2 \ \sum_{i=1}^k \sigma_i^2(W_1)\sigma_i^2(W_2) \ < \     \alpha_k^2 k
    \end{equation}
    then~\eqref{eq:net_sys} is $k$-contracting  w.r.t. the $L_2$ norm. 
    Furthermore, if these conditions hold for~$k=2$ then every bounded trajectory of~\eqref{eq:net_sys} converges to an equilibrium point (which is not necessarily unique).
\end{Theorem}

\begin{Remark}
    The condition~$\alpha_k>0$ amounts  to requiring that for any~$x\in\Omega$ 
the sum of every~$k$ eigenvalues of the matrix
\begin{equation}\label{eq:D_net}
D
(x)
:=\diag(d'_1(x_1),\dots,d'_n(x_n))
\end{equation}
is positive. In particular, for~$k=1$, this amounts to requiring that~$D(x)$ is a positive diagonal matrix for any~$x\in\Omega$, but for~$k>1$ some of the~$d_i'$s may be negative, as long as the sum of every~$k$ of the~$d_i'$s is positive.

It is natural to expect that a sufficient condition for~$k$-contraction   implies~$j$-contraction for any~$j>k$. Note that if~$a_i \in\R$ with~$a_1\geq\dots \geq a_n$, and~$k\in\{1,\dots,n-1\}$ then 
\[
        \frac{1}{k}\sum_{i=1} ^k a_i - \frac{1}{k+1}\sum_{i=1} ^{k+1} a_i \geq \frac{  a_k-a_{k+1}}{k+1}\geq 0.
\]
Using this 
it is straightforward to verify that if~\eqref{cond:net_k_smallgain} holds for some~$k$ then it also holds for~$k+1$, and thus for any~$j>k$. 

\end{Remark}

\begin{IEEEproof}[Proof of Thm.~\ref{thm:net_k_contract}]
By~\eqref{cond:net_k_smallgain}, there exists~$\gamma > 0$ such that
\be\label{eq:gammprop}
    \gamma < \alpha_k \text{ and } {\sup_{x \in \Omega}} \|J_f(W_2x)\|_2^2 \ \sum_{i=1}^k \sigma_i^2(W_1) \sigma_i^2(W_2)< \gamma^2 k.
\ee
Represent~\eqref{eq:net_sys} as 
a GLS, namely, the interconnection of the nonlinear  system   
\begin{align}\label{eq:ygxx}
    \dot x  = -d(x) + v + \gamma u,  \quad 
    y =x, 
\end{align}
and the  time-invariant  nonlinearity
\be\label{eq:tinon}
\Phi(y) := -\gamma^{-1}W_1f(W_2 y),
\ee
where the input to the dynamical system is  $u = \Phi(y)$.
We now apply  Thm.~\ref{thm:suff} 
to  this~GLS. Let~$\Theta = q I$,  with~$q > 0$. Then~\eqref{cond:diff_ARI_eig} becomes
\begin{equation}\label{eq:diff_ARI_eig_net}
    -2\sum_{i=1}^k \lambda_i(D(x)) + k q^2 \gamma^2 + k q^{-2} \le -\eta_1,
\end{equation}
with~$D$ as in~\eqref{eq:D_net}.
%
By \eqref{eq:alphak},~$ k\alpha_k \le \sum_{i=1}^k \lambda_i(D(x)) $,
so~\eqref{eq:diff_ARI_eig_net} will hold if
$
\left( -2 \alpha_k + \gamma^2 p + p^{-1} \right) k \leq -\eta_1,
$  
with~$p :  = q^2$.
Taking~$p:=\alpha_k\gamma^{-2}$ gives that~\eqref{eq:diff_ARI_eig_net} indeed  holds for~$\eta_1= (\alpha_k^2-\gamma^{2})  / \alpha_k$, and~\eqref{eq:gammprop} implies that~$\eta_1>0$.

To verify that condition~\eqref{cond:C_JPhi}
in Thm.~\ref{thm:suff}
also holds, note that in~\eqref{eq:ygxx} we have~$y=g(x)=x$, so we can use the approach in Remark~\ref{rem:C_is_I}. 
Using Lemma~\ref{lemmma:sing}  
and~\eqref{eq:gammprop} gives
\begin{align*}
    \sum_{i=1}^k \sigma_i^2(J_\Phi)
& = \sum_{i=1}^k \sigma_i^2 (-\gamma^{-1} W_1 J_f W_2)\\
&\leq\gamma^{-2}  \sum_{i=1}^k \sigma_i^2 (  W_1 J_f )  \sigma_i^2 (  W_2  ) \\
&    \leq \gamma^{-2} \sigma_1^2 (    J_f )  \sum_{i=1}^k \sigma_i^2 (  W_1   )  \sigma_i^2 (  W_2  ) \\
&<k.
\end{align*}
Thus, 
there exists~$\eta_2>0$
such that~\eqref{eq:pm2} holds,
and Thm.~\ref{thm:suff} implies that~\eqref{eq:net_sys} is $k$-contracting. Since $P=pI_n$,  the closed-loop system
is $k$-contracting  w.r.t. the (unweighted) $L_2$ norm.

If~\eqref{cond:net_k_smallgain} holds with~$k=2$ then~\eqref{eq:net_sys} is a time-invariant 2-contracting system and  
this implies  that every bounded trajectory converges to an equilibrium point~\cite{li1995}. This completes the proof of Thm.~\ref{thm:net_k_contract}.
\end{IEEEproof}

\begin{Remark}
In the context of neural network models, several  approaches for stability analysis require that the connection matrix (or matrices) is symmetric~\cite{cohen_NN, russo_symm}.  Thm.~\ref{thm:net_k_contract}
includes no such assumption.
 \end{Remark}

\begin{Remark}\label{rem:spect}
Let~$\xi^1,\dots,\xi^n$ denote the canonical basis in~$\R^n$. 
    Suppose that in  the networked system~\eqref{eq:net_sys}
    we have~$q=1$ and~$W_2=c (\xi^i)^\T$ for some~$c\in\R$ and~$i\in[1,n]$.
    Then~\eqref{eq:tinon} becomes 
  $
\Phi(y) = -\gamma^{-1}W_1f(c x_i), 
$
that is, the representation of the network as a GLS is
``single-output'' in the sense that the nonlinear feedback depends only on the state-variable~$x_i$. The singular values of~$W_2$ are~$|c|,0,\dots,0$, 
so the sufficient 
condition in~\eqref{cond:net_k_smallgain}
    simplifies to 
$
         {\sup_{x \in \Omega}} \|J_f(c x_i)\|_2^2 \ c^2 \sigma_1^2(W_1) \ < \     \alpha_k^2 k. 
  $ 
  A similar simplification occurs if the GLS is
``single-input'', i.e.,~$m=1$ and~$W_1=c \xi^i$ for some~$c\in\R$
and~$i\in[1,n]$. 
\end{Remark}

\subsection{A biochemical control circuit}
Consider the feedback system
\begin{align}\label{eq:smith_feedback}
    \dot x_1 &= -d_1(x_1)+r(x_n),\nonumber\\
    \dot x_2 &= -d_2(x_2)+x_1,\nonumber\\
    &\vdots\\
    \dot x_n &= -d_n(x_n)+x_{n-1},\nonumber
\end{align}
with~$d_i,r:\R\to\R$ continuously differentiable. 

\begin{Remark}
    Ref.~\cite[Chap.~4]{hlsmith} used this model, in the particular
case of linear dissipation, that is, 
$
d_i(x_i)=c_i x_i$, $c_i>0$,
and the positive feedback function
$
r(s)=(1+s)/(\zeta+s)$,
with~$  \zeta>1$, 
as a model of the 
control of protein synthesis in the cell. Every~$x_i$ represents the concentration of mRNA molecules or enzymes or products, so the state space is~$\Omega=\R^n_+$. 
This system was also analyzed   using contraction theory in~\cite{weak_contraction}. 
\end{Remark}

We   analyze the more general case of not necessarily linear dissipation and a general feedback function~$r$ using Thm.~\ref{thm:net_k_contract}. Write~\eqref{eq:smith_feedback} as the networked system~\eqref{eq:net_sys} 
with~$d(x)=\begin{bmatrix} d_1(x_1) & \dots & d_n (x_n) \end{bmatrix}^\T $, $W_1 = W_2 = I_n$, $v=0$, and~$f(y)=\begin{bmatrix} r(y_n) & y_1 & \dots & y_{n-1} \end{bmatrix}^\T$. Thus, 
$
\|J_f(W_2 x)\|_2^2 = \max\{(r'(x_n))^2, 1\}. 
$
The next result follows   from Thm.~\ref{thm:net_k_contract}.
\begin{Corollary}\label{coro:sm}
Assume that~$\Omega = \R^n_+$ is an invariant set of~\eqref{eq:smith_feedback}. Fix~$k\in[1,n]$. Suppose that~$\alpha_k>1$,
and that
\be\label{eq:cond2_sm}
     \max_{x_n\geq 0} (r'(x_n))^2  < \alpha_k^2 . 
\ee
Then~\eqref{eq:smith_feedback}  is~$k$-contracting on~$\Omega$. 
\end{Corollary}

Corollary~\ref{coro:sm} provides a simple and easy to check sufficient condition for~$k$-contraction for any~$k\in[1,n]$.
Note that this condition depends on a balance between the ``$k$-total dissipation''~$\alpha_k$ on the one hand,
and the maximal derivative of the feedback function~$r$ on the other hand. Also,  the analysis in~\cite[Chapter 4]{hlsmith} is based on the theory of irreducible cooperative systems, and this requires that~$r'(s)\geq 0$ for all~$s\geq 0$. Corollary~\ref{coro:sm} does not  constrain  the sign of~$r'$. In particular,~$r'(s)$ may have a different sign for different values of~$s$.

As a specific   example, consider~\eqref{eq:smith_feedback} with~$n=3$, $d_1(x_1) = \sin(x_1) + (1/2)$, $d_2(x_2)= 3 x_2$, $d_3(x_3)=3 x_3$, and~$r(s)=(1+s)/(2+s)$, that is,
\begin{align}\label{eq:exa_sm} 
    \dot x_1 &= - \sin(x_1) - \frac{1}{2} + \frac{1+x_3}{2+x_3},\nonumber\\
    \dot x_2 &= -3x_2 + x_1, \\
    \dot x_3 &= -3x_3 + x_2.\nonumber
\end{align}
Since~$r(s) \geq 1/2$ for all~$s \geq 0$, we have that~$\dot x_1\geq -\sin(x_1)$, implying  that~$\R^3_+$ is an invariant set of~$\eqref{eq:exa_sm}$.

\begin{figure}
    \centering
    \includegraphics[scale=0.4]{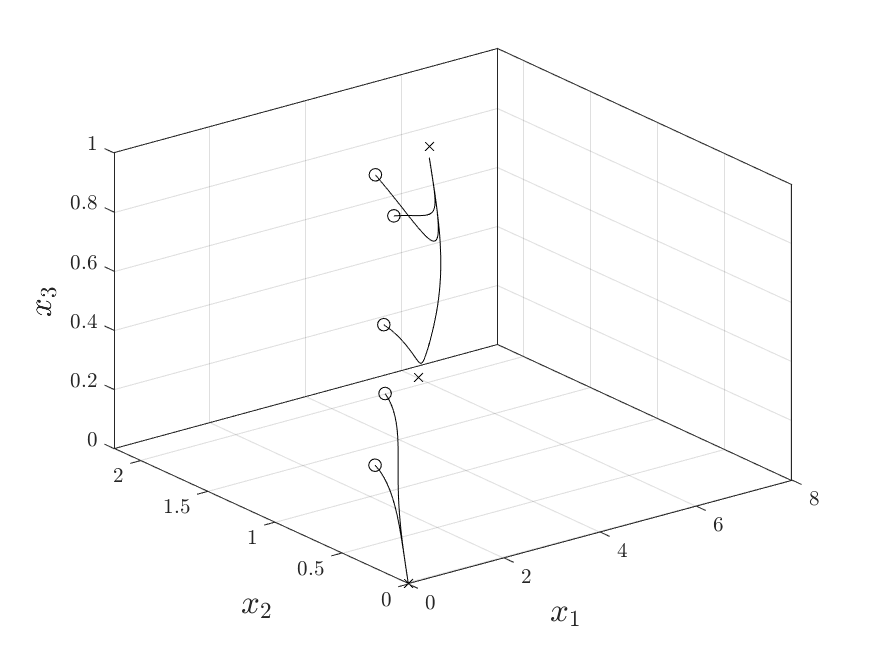}
    \caption{Trajectories of~\eqref{eq:exa_sm} from five   initial conditions marked by circles. Several equilibrium points are marked by a cross.}
    \label{fig:five}
\end{figure}

A point~$e\in\R^3_+$ is an equilibrium  of~\eqref{eq:exa_sm} if $e_2= 3e_3$, $e_1 = 3e_2=9e_3$,
and
$
 \sin(9e_3) + \frac{1}{2} = \frac{1+e_3}{2+e_3}. 
$
It is straightforward to verify that there are infinitely many equilibrium points in~$\R^3_+$,
so the system is not contracting (i.e., not~$1$-contracting) w.r.t. any norm. 
Consider the conditions in Corollary~\ref{coro:sm} for~$k=2$. Here,
$
\alpha_2=3/2 > 1,
$
and
$
 r'(x_3)=(2+x_3)^{-2}, 
$  so~$r'(x_3)\leq 1/4$ for all~$x_3\geq 0$, and~\eqref{eq:cond2_sm} holds for~$k=2$.  
We conclude that the system is~$2$-contracting, so every bounded solution converges to an equilibrium. Fig.~\ref{fig:five} depicts three equilibrium points and trajectories
of~\eqref{eq:exa_sm}  for five (randomly chosen) initial conditions in~$\R^3_+$. As expected, every solution converges to an equilibrium.

\section{Conclusion}
A GLS is 
the feedback connection of nonlinear dynamical system and a nonlinear feedback function.  GLSs typically have more than a single equilibrium and are
thus not~$1$-contracting w.r.t. any norm. 
If they are $2$-contracting  then 
every bounded solution converges to an equilibrium (which is not necessarily unique) 
and establishing such a global property is important in many applications. For example, in a neural network model of an associative memory every stored memory pattern  corresponds to an equilibrium, and $2$-contraction implies that for any initial condition 
the solution will indeed converge to some  stored pattern. 

Topics for further research include: (1)~a given  nonlinear system can often be represented  as a GLS in several ways. How can one  find the ``best'' possible representation? 
(2)~our results only consider $k$-contraction w.r.t. (scaled) $L_2$ norms and it may be of interest to analyze  $k$-contraction w.r.t. other norms; and 
(3)~Thm.~\ref{thm:net_k_contract}
provides a sufficient condition for $k$-contraction that  uses the singular values of the  matrices~$W_1$,   $W_2 $ in the networked system~\eqref{eq:net_sys}. 
In other words, it is based on the spectral properties of the connection graphs (see also Remark~\ref{rem:spect}). 
An interesting research  direction  is to  further develop this into a graph-theoretic approach to~$k$-contraction.  \\
\emph{Acknowledgments.}
We thank the editor and the anonymous reviewers
for many helpful and detailed comments.

\vspace*{-0.3cm}
\bibliographystyle{IEEEtranS}
\bibliography{fullrefs,refs}

\end{document}